\documentclass[12pt]{iopart}

\begin{document}

\title[Non-Noether symmetries and bidifferential calculi]{Remark on
non-Noether symmetries and bidifferential calculi}
\author{George Chavchanidze}
\address{Department of Theoretical Physics, 
A. Razmadze Institute of Mathematics,\\ 
1 Aleksidze Street, Ge 380093, 
Tbilisi, Georgia\\
e-mail: chaffinch@apexmail.com}

\begin{abstract}
In this paper relationship between non-Noether symmetries and
bidifferential calculi is 
disscussed. 
\end{abstract}

 In the past few years both non-Noether symmetries [1-4] and
bidifferential calculi [5-8]
has been successfully used in generating conservation laws and both lead
to the similar
families of conserved quantities. \\
Here relationship between Lutzky's integrals of motion [3-4] and
bidifferential calculi
is briefly disscussed. \\
In the regular Hamiltonian system phase space M is equipped with
symplectic form $\omega$
(closed $d\omega=0$ and non-degenerate $i_X\omega =0 \Leftrightarrow X=0$
2-form) which gives rise to 
the isomorphism between the vector fields and 1-forms 
\begin{equation} \phi_\omega: TM \longrightarrow T^*M
\end{equation}
defined by
\begin{equation}
\phi_\omega(X)=i_X\omega
\end{equation}
where $i_X\omega$ denotes contraction of the vector field $X$ with
$\omega$. This map can be
inverted
\begin{equation} \phi_\omega^{-1}: T^*M \longrightarrow TM
\end{equation}
with $\phi_\omega^{-1}(a)=i_Wa~~a\in \Omega^1(M)$ where $W$ is a bivector
field obtained by 
inverting $\omega$. The vector field is said to be Hamiltonian (locally
Hamiltonian)
if it corresponds to exact (closed) form 
\begin{equation} X_f=\phi_\omega^{-1}(df) \end{equation}
According to Liouville's theorem  Hamiltonian vector field 
preserves both $\omega$ ($L_{X_f}\omega=0$ where $L_{X_f}$ denotes Lie
derivative along $X_f$ vector field)
and bivector field $W$ ($[X_f,W]=0$ where $[~,~]$ is a Schoulten
bracket).\\
Now let's consider non-Hamiltonian vector field $E$ ($L_E\omega \neq
0~~~[E,W] \neq 0$) commuting
with the tangent vector of solutions
$[E,X_h]=0~~X_h=\phi_\omega^{-1}(dh)$. The infinitesimal transformation
$g=1+\epsilon L_E$ generated by $E$ is a non-Noether symmetry [3-4] in
sense that it doesn't preserve 
action but maps space of solutions onto itself. \\
As it was shown by Lutzky [3] such a symmetry leads to the whole family of
conserved quantities
\begin{equation}l_k = i_{W^k}\omega_E^k ~~~~ k=1,2...n \end{equation}
\begin{center}(no summation over k)\end{center}
where $\omega_E=L_E\omega$ and $n$ is dimension of configuration space. \\
Indeed 
\begin{eqnarray}
\frac{d}{dt}l_k=L_{X_h}l_k=L_{X_h}i_{W^k}\omega_E^k=i_{[X_h,W^k]}\omega_E^k+i_{W^k}L_{X_h}\omega_E^k=0
\end{eqnarray}
since $[X_h,W]=L_{X_h}\omega =0$ (according to the Liouville's theorem)
and 
$L_{X_h}\omega_E=L_{X_h}L_E\omega=L_EL_{X_h}\omega=0$ (due to
$[X_h,E]=0$). \\
The whole construction can be extended to the singular dynamical systems
and to the more 
general class of symmetries [4].\\
Now let's turn to the bidifferential calculi. The possibility of
generating conservation laws 
by means of exterior derivative $d$ and auxiliary differential operator
$d'$ with properties
$d'^2=0$ and $dd'+d'd=0$ has been considered by Dimakis and M\"{u}ller
-Hoissen [7-8].\\
According  to the Fr\"{o}licher-Nijenhuis theory action of $d'$ on
functions should be of 
the following form: $d'f=i_Adf$ where $A\in \Omega^1(M,TM)$ is tangent
valued 1-form 
with vanishing Fr\"{o}licher-Nijenhius torsion [5,6]. 
\begin{eqnarray}
T(A)(X,Y)=[AX,AY]-A([AX,Y]+[X,AY]-A[X,Y])=0 \nonumber \\
T(A)\in \Omega^2(M,TM)~~X,Y\in TM 
\end{eqnarray}
One can show [5] that traces of powers 
of $A$ 
\begin{equation}
\mu_k=Tr(A^k)
\end{equation}
form a Lenard scheme 
\begin{equation}
d\mu_{k+1}=d'\mu_k
\end{equation}
and as a result $\{\mu_k\}$ are in involution - have vanishing Poisson
brackets 
$\{\mu_k,\mu_{k'}\}=0$ [6]. This sequence of functions in involution
associated with 
bidifferential  calculi $d,d'$ can be used in generating conservation
laws. \\
Now let's expose relationship between the family of functions $\{\mu_k\}$
mentioned above 
and conserved quantities associated with non-Noether symmetry. \\
By means of the generator of symmetry $E$ one can construct differential
operator $\hat d$
with all properties of $d'$ besides $\hat d^2=0$. The action of $\hat d$
on functions can 
be defined by 
\begin{equation}
\hat df=\phi_{\omega_E}\circ\phi_\omega^{-1}(df)=\hat A(f)
\end{equation}
where $\hat A=(\omega_E)_{ik}W^{kj}dx_i\otimes \frac{\partial}{\partial
x_j} $.
Since $W$ and $\omega_E$ are invariants ($L_{X_h}\omega=L_{X_h}W=0$
according to Liouville's theorem
and $L_{X_h}\omega_E=L_{X_h}L_E\omega=L_EL_{X_h}\omega=0$) therefore
operator 
\begin{equation}
\phi_{\omega_E}\circ\phi_\omega^{-1}:T^*M\longrightarrow T^*M
\end{equation}
is also constant along solutions and as a result the eigenvalues of 
$\phi_{\omega_E}\circ\phi_\omega^{-1}$ 
are integrals of motion. \\
So if $\{\lambda_k\}$ are eigenvalues of
$\phi_{\omega_E}\circ\phi_\omega^{-1}$ it is easy to verify that
Lutzky's conserved quantities $\{l_k\}$ are expressed by means of
$\{\lambda_k\}$
\begin{equation}
l_k=\sum_{i_s \neq i_{s'}}\lambda_{i_1}\lambda_{i_2}...\lambda_{i_k}
\end{equation} 
and vice versa. Traces of powers of $\hat A$ are combinations of
$\{\lambda_k \}$ too
\begin{equation}
\hat \mu_k=Tr(\hat A^k)=\sum_i \lambda_i^k
\end{equation}
and therefore are also integrals of motion associated with non-Noether
symmetry generated 
by $E$ ($\{\hat \mu_k\}$ can be expressed by means of $\{l_k\}$ and vice
versa).\\
In general we have no restriction on torsion of $\hat A$ (this reflects
the fact that in 
general Lutzky's conserved quantities are not in involution) but when it
vanishes $\{\hat \mu_k\}$
reduces to the sequence of functions $\{\mu_k\}$
\begin{equation}
d\mu_{k+1}=\hat d\mu_k
\end{equation}
associated with bidifferential calculi $d,\hat d$ and considered by
Dimakis and M\"{u}ller-Hoissen. \\
\ack{Author is  indebted to Z. Giunashvili for many useful discussions 
and G. Jorjadze for invaluable help. This work was supported by
Scholarship from World 
Federation of Scientists.}

\end{document}